\documentclass[twocolumn,showpacs,preprintnumbers,amsmath,amssymb]{revtex4}
\usepackage{graphicx}
\usepackage{textcomp}

\begin{document}
\draft
\title{Field-induced decay of quantum vacuum: visualizing  pair production in a classical photonic system}
\normalsize

\author{S. Longhi}
\address{Dipartimento di Fisica, Politecnico di Milano, Piazza L. da Vinci
32, I-20133 Milano, Italy}

%\date{.}

%
\bigskip
\begin{abstract}
\noindent The phenomenon of vacuum decay, i.e. electron-positron
pair production due to the instability of the quantum
electrodynamics vacuum in an external field, is a remarkable
prediction of Dirac theory whose experimental observation is still
lacking. Here a classic wave optics analogue of vacuum decay, based
on light propagation in curved waveguide superlattices, is proposed.
Our photonic analogue enables a simple and experimentally-accessible
visualization in space of the process of pair production as break up
of an initially negative-energy Gaussian wave packet, representing
an electron in the Dirac sea, under the influence of an oscillating
electric field.
\end{abstract}


%42.79.Gn    Optical waveguides and couplers
%78.67.Pt    Multilayers; superlattices; photonic structures; metamaterials
%03.65.Pm    Relativistic wave equations

\maketitle

\section{Introduction}
Quantum-classical analogies have been exploited on many occasions to
mimic at a macroscopic level many quantum phenomena which are
currently inaccessible in microscopic quantum systems
\cite{Dragomanbook}. In particular, in the past decade engineered
photonic structures have provided a useful laboratory tool to
investigate and visualize with classical optics the dynamical
aspects embodied in a wide variety of coherent quantum phenomena
encountered in atomic, molecular, condensed-matter and matter-wave
physics \cite{Longhi09LPR}. Among others, we mention the optical
analogues of electronic Bloch oscillations \cite{B01,B02} and Zener
tunneling \cite{ZT,Dreisow09}, dynamic localization \cite{DL},
coherent enhancement and destruction of tunneling \cite{CDT},
adiabatic stabilization of atoms in ultrastrong laser fields
\cite{stabi}, Anderson localization \cite{AL}, quantum Zeno effect
\cite{QZ}, coherent population transfer \cite{STIRAP}, and coherent
vibrational dynamics \cite{molecule}. Most of the above mentioned
studies are based on the formal analogy between paraxial wave
equation of optics in dielectric media and the single-particle
nonrelativistic Schr\"{o}dinger equation, thus providing a test bed
for classical analogue studies of nonrelativistic quantum phenomena.
Recently, a great attention has been devoted toward the
investigation of experimentally-accessible and controllable
classical or quantum systems that simulate certain fundamental
phenomena rooted in the relativistic Dirac equation. Among others,
cold trapped atoms, ions and graphene have proven to provide
accessible systems to simulate relativistic physics in the lab, and
a vast literature on this subject has appeared in the past few years
(see, for instance, \cite{GR1,GR2,ion} and references therein). In
particular, low-energy excitations of nonrelativistic
two-dimensional electrons
 in graphene obey
the Dirac-Weyl equation and behave like massless relativistic
particles. This has lead to the predictions in condensed-matter
physics of phenomena analogous to Zitterbewegung
\cite{Zitterbewegung} and Klein tunneling \cite{Klein} of
relativistic massive or massless particles, with the first
experimental evidences of Klein tunneling in graphene \cite{EGR1}
and in carbon nanotube \cite{EGR2} systems.
 Photonic analogues of Dirac-type equations have been also theoretically proposed
for light propagation in certain triangular or honeycomb photonic
crystals, which mimic conical singularity of energy bands of
graphene \cite{Haldane,Zhang08}, as well as in metamaterials
\cite{meta} and in optical superlattices \cite{Longhi09un}. This has
leads to the proposals of photonic analogues of relativistic
phenomena like Zitterbewegung \cite{Zhang08,meta,Longhi09un} and
Klein tunneling \cite{Segev09,meta}.\\
Electron-positron pair production due to the instability of the
quantum electrodynamics (QED) vacuum in an external electric field
(a phenomenon generally referred to as vacuum decay) is another
remarkable prediction of Dirac theory and regarded as one of the
most intriguing non-linear phenomena in QED (see, for instance,
\cite{Fradkin,Avetissian}). In intuitive terms and in the framework
of the one-particle Dirac theory, the pair production process can be
simply viewed as the transition of an electron of the Dirac sea
occupying a negative-energy state into a final positive-energy
state, leaving a vacancy (positron) in the negative-energy states.
There are basically two distinct transition mechanisms: the
Schwinger mechanism in presence of an ultrastrong static electric
field, and dynamic pair creation in presence of time-varying
electric fields. The Schwinger mechanism \cite{Schwinger51}  can be
understood as a tunneling process through a classically forbidden
region, bearing a close connection to Klein tunneling. Dynamic pair
creation, originally proposed by Brezin and Itzykson \cite{Brezin70}
for oscillating spatially-homogeneous fields and subsequently
investigated by several authors (see, for instance,
\cite{oscillating1,oscillating2,oscillating3} and references
therein), has attracted recently great attention since the advent of
ultrastrong laser system facilities, which pave the way toward the
realization of purely laser-induced pair production \cite{note1}.
Electric fields oscillating in time only can be achieved at the
antinodes of two oppositely propagating laser beams, and can lead to
such intriguing phenomena as Rabi oscillations of the Dirac sea
(see, for instance, \cite{Avetissian,oscillating2}). In the
framework of the one-particle Dirac theory of vacuum decay
\cite{oscillating2}, a simple picture of pair production is
represented by the time evolution of an initially negative-energy
Gaussian wave packet, representing an electron in the Dirac sea,
under the influence of an oscillating electric field
\cite{oscillating3}. When the $e^+e^-$ pair is produced, a droplet
is separated from the wave packet and moves opposite to the initial
one \cite{oscillating3}. The droplet is a positive-energy state and
represents the created electron. Such a dynamical scenario of pair
production, observed in numerical simulations \cite{oscillating3},
is unlikely to be accessible in a foreseen experiment using two
counterpropagating and ultraintense laser beams, therefore it may be
of some interest to find either a classical or a quantum simulator
capable of visualizing in the lab the wave packet dynamics of
$e^+e^-$ pair creation.  It is the aim of this work to propose a
classic wave optics analogue of the QED $e^+e^-$ pair production in
oscillating fields, based on monochromatic light propagation in
curved waveguide optical superlattices \cite{Dreisow09,Longhi06OL},
in which the temporal evolution of Dirac wave packets and $e^+e^-$
pair production is simply visualized as spatial beam break up along
the curved photonic structure. As our analysis is focused onto a
classical wave optics analogue of QED vacuum decay, a similar
dynamical scenario could occur for ultracold atoms in accelerated
double-periodic optical lattices (see, for instance, \cite{Breid}),
which might thus provide a quantum simulator for relativistic QED
decay complementary to our classical analogue.

\section{Basic model and quantum-optical analogy}

\subsection{The photonic structure}
The photonic structure considered in this work consists of a
one-dimensional binary superlattice of weakly-guiding dielectric
optical waveguides with a periodically-curved optical axis. This
optical system was previously introduced in another physical context
for multiband diffraction and refraction control of light waves in
optical lattices \cite{Longhi06OL}. It should be noted that a
similar photonic structure, but with a circularly-curved optical
axis, has been recently realized to visualize in the lab the optical
analogue of coherent Bloch-Zener oscillations of nonrelativistic
electrons in crystalline potentials \cite{Dreisow09}, whereas the
same structure but with a straight optical axis has been proposed to
realize a photonic analogue of Zitterbewegung and Klein tunneling of
relativistic electrons \cite{Longhi09un}. The photonic superlattice
consists of two interleaved and equally spaced lattices A and B, as
shown in Fig.1(a), with spacing $a$ between adjacent waveguides. The
normalized waveguide index profile is assumed to be the same for the
waveguides of the two lattices A and B, with an alternation of peak
refractive index changes $dn_1$ and $dn_2$ for the two lattices
\cite{Dreisow09}. The waveguide axis is periodically bent along the
propagation direction $z$ [see Fig.1(b)], with an axis bending
profile $x_0(z)$ of period $\Lambda$; typically the condition
$\dot{x}(0)=0$ is assumed to ensure that the waveguide axis is
orthogonal to the exciting input plane $z=0$, where the dot denotes
the derivative with respect to $z$. Light propagation in the curved
superlattice is at best captured in the 'waveguide' reference frame
after a Kramers-Henneberger transformation, where the array appears
to be straight (see, for instance, \cite{Longhi09LPR}). In this
reference frame, the waveguide curvature along the propagation axis
appears as a fictitious refractive index gradient in the transverse
plane with a $z$-varying slope proportional to the local waveguide
axis curvature \cite{Longhi09LPR}. After a gauge transformation for
the electric field envelope, propagation of monochromatic waves at
wavelength (in vacuum) $\lambda$ is described by the following
scalar wave equation (see, for instance,
\cite{Longhi09LPR,Longhi06OL})
\begin{equation}
i \frac{\partial E}{\partial z} = -\frac{\lambda}{4 \pi n_s}
\frac{\partial^2 E}{\partial x^2} + 2 \pi \frac{n_s-n(x)}{\lambda}
E+\frac{2 \pi n_s \ddot{x}_0(z)}{\lambda} x E
 \end{equation}
where $n_s$ is the bulk refractive index, $n(x)$ is the refractive
index profile of the superlattice, and the double dot stands for the
second derivative with respect to $z$. In the tight-binding
approximation, light transport in the curved photonic lattice can be
described by means of coupled-mode equations for the amplitudes
$c_l$ of light modes trapped in the various waveguides
\cite{Longhi06OL}
\begin{equation}
i \frac{dc_l}{dz}=-\sigma (c_{l+1}+c_{l-1})+(-1)^l \delta c_l+ F(z)
l c_l,
\end{equation}
where $2 \delta$ and $\sigma$ are the propagation constant mismatch
and the coupling rate between two adjacent waveguides of lattices A
and B, and $F(z)=2 \pi n_s a \ddot{x}_0(z)/\lambda$. The
tight-binding model (2) is accurate to describe beam dynamics of
Eq.(1) provided that the first two minibands of the array are
involved in the dynamics \cite{Longhi06OL}. A similar tight-binding
model also applies to the dynamics of cold atoms in accelerated
double-periodic optical lattices \cite{Breid}. For a straight
lattice, i.e. for $F(z)=0$, the tight-binding minibands of the
superlattice are readily calculated by making the plane-wave Ansatz
$c_l(q) \sim \exp(iqla-i\omega z)$ in Eq.(2) and read [see Fig.1(c)]
\begin{equation}
\omega_{\pm}(q)= \pm \sqrt{\delta^2+4 \sigma^2 \cos^2 (qa)}
\end{equation}
\begin{figure}[htbp]
\includegraphics[width=85mm]{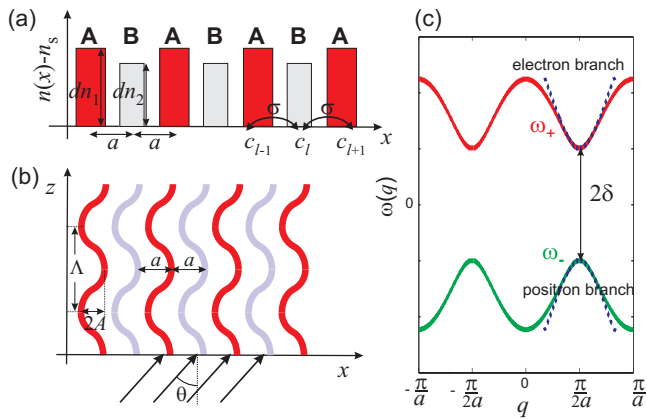}\\
   \caption{(color online) (a) and (b) Schematic of a binary superlattice, made of two interleaved lattices A and B of waveguides with
   high ($dn_1$) and low ($dn_2$) refractive index changes, equally spaced by $a$ [Fig.1(a)] and with a periodically-curved axis [Fig.1(b)].
The array is excited by  a broad beam tilted at the angle $\theta$.
(c) Dispersion curves (minibands) of the tight-binding binary
superlattice [Eq.(3)] for the straight binary array (solid curves),
and corresponding electron and positron dispersion curves of the
Dirac equation (7) (in absence of the ac field) near the Brillouin
zone edge $q=\pi/(2a)$ (dotted curves).}
\end{figure}
where $q$ is the wave number.

\subsection{Quantum-optical analogy}
In Ref.\cite{Longhi09un}, it was recently shown that, near the
Brillouin zone edge $q= \pm \pi/(2a)$, the dispersion relations (3)
of the straight superlattice approximate the positive (electron) and
negative (positron) energy curves of a one-dimensional massive Dirac
electron in absence of external fields, and that beam propagation at
incidence angles near the Bragg angle mimics the temporal dynamics
of a one-dimensional free Dirac electron, the two components of the
spinor wave function corresponding to the occupation amplitudes in
the two sublattices A and B of the superlattice. To study the
photonic analogue of vacuum decay and electron-positron production
arising from an oscillating field, we consider here light
propagation in a periodically-curved binary array within the
tight-binding model (2). As it will be shown below, the undulation
of the waveguide axis along the propagation direction mimics an
external ac electric field in the one-dimensional Dirac equation. To
explain the quantum-optical analogy, let us set in Eqs.(2)
$c_l(z)=a_l(z) \exp[-i \Phi(z)l]$, where
\begin{equation}
\Phi(z)=\int_0^z d \xi F(\xi) =\frac{2 \pi n_s a
\dot{x}_0(z)}{\lambda}.
\end{equation}
Coupled-mode equations than take the equivalent form
\begin{eqnarray}
i \frac{da_l}{dz} & = & -\sigma \exp[-i \Phi(z)]a_{l+1}- \sigma
\exp[i \Phi(z)]a_{l-1} \nonumber \\
& + &(-1)^l \delta a_l.
\end{eqnarray}
Note that, for the special case of a sinusoidal axis bending profile
$x_0(z)=-A \cos( 2 \pi z / \Lambda)$ of amplitude $A$ and period
$\Lambda$, one has $\Phi(z)=\Phi_0 \sin(2 \pi z / \Lambda)$ with
\begin{equation}
\Phi_0=\frac{4 \pi^2 n_s a A}{\lambda \Lambda}.
\end{equation}
 Let us then introduce the following two
assumptions: (i) the amplitude of axis bending profile is small
enough that $|\Phi| \ll \pi/2$ over the entire oscillation cycle;
(ii) the array is excited by a broad beam tilted at the Bragg angle
$\theta_B=\lambda/(4n_sa)$. The first assumption enables to set
$\exp( \pm i \Phi) \sim 1 \pm i \Phi$ in Eqs.(5); the second
assumption implies that the modes in adjacent waveguides are excited
with a nearly equal amplitude but with a phase difference of $\pi
/2$ \cite{Longhi09un}. Therefore, after setting
 $a_{2l}(z)=(-1)^l \psi_1(l,z)$, $a_{2l-1}=-i(-1)^l \psi_2(l,z)$,
the amplitudes $\psi_1$ and $\psi_2$ vary slowly with $l$, and one
can thus write $\psi_{1,2}(l \pm 1,z)=\psi_{1,2}(l ,z) \pm (\partial
\psi_{1,2}/\partial l)$ and consider $l \equiv \xi=x/(2a)$ as a
continuous variable rather than as an integer index.  Under such
assumptions, from Eqs.(5) it readily follows that the two-component
spinor $\psi(\xi,z)=(\psi_1,\psi_2)^T$ satisfies the one-dimensional
Dirac equation in presence of an external ac field, namely
\begin{equation}
i \frac{\partial \psi}{\partial z}= -i \sigma \alpha \frac{\partial
\psi}{\partial \xi}-2 \sigma \Phi(z) \alpha \psi + \delta \beta
\psi,
\end{equation}
where
\begin{equation}
\alpha=\left(
\begin{array}{cc}
0 & 1 \\
1 & 0
\end{array}
\right) \; , \; \beta=\left(
\begin{array}{cc}
1 & 0 \\
0 & -1
\end{array}
\right)
\end{equation}
are the $\sigma_x$ and $\sigma_z$ Pauli matrices, respectively. Note
that, after the formal change
\begin{eqnarray}
\sigma & \rightarrow & c \nonumber \\
 \delta & \rightarrow & \frac{ mc^2}{
\hbar} \nonumber \\
\xi & \rightarrow & x \\
\Phi & \rightarrow & \frac{eA_x} {2 \hbar c} \nonumber \\
z & \rightarrow & t, \nonumber
\end{eqnarray}
 Eq.(7) corresponds to the one-dimensional Dirac
equation for an electron of mass $m$ and charge $e$ in presence of a
spatially-homogeneous and time-varying vectorial potential
$\mathbf{A}=(A_x,0,0)$, which describes the interaction of the
electron with an external oscillating electric field
$E_x(t)=-(\partial A_x /
\partial t) $ in the dipole approximation (see, for instance, \cite{book}). Note also that, in our optical
analogue, the {\it temporal} evolution of the spinor wave function
$\psi$ for the Dirac electron is mapped into the {\it spatial}
evolution of the field amplitudes $\psi_1$ and $\psi_2$ along the
$z$-axis of the array, and that the two components $\psi_1$,
$\psi_2$ of the spinor wave function correspond to the occupation
amplitudes in the two sublattices A and B composing the
superlattice. In absence of the external field, the energy-momentum
dispersion relation $\hbar \omega(k)$ of the Dirac equation (7),
obtained by making the Ansatz $\psi \sim \exp(ik \xi-i \omega z)$,
is composed by the two branches $ \omega_{\pm}(k)= \pm \epsilon(k)$,
corresponding to positive and negative energy states of the
relativistic free electron, where
\begin{equation}
\epsilon(k)=\sqrt{\delta^2+\sigma^2 k^2}.
\end{equation}
 Note that such two
branch curves are readily obtained from Eq.(3) after setting
\begin{equation}
q= \frac{\pi}{2a}+\frac{k}{2a}
\end{equation}
 and assuming small values of $k$, which corresponds to
 the assumption of slowly-varying envelopes $\psi_{1,2}(l,z)$.
Therefore, the positive and negative energy states of the free Dirac
electron are mapped into the dispersion curves of the two minibands
of the binary array near the boundary of the Brillouin zone at
$q=\pi/(2a)$, as shown in Fig.1(c).

\section{Optical analogue of pair production in an oscillating field}
The phenomenon of vacuum decay by oscillating fields, originally
predicted in Ref.\cite{Brezin70}, refers to the electron-positron
pair production due to the instability of the QED vacuum in a
superstrong field of two counterpropagating laser beams
\cite{oscillating2,oscillating3}. Vacuum decay should be generally
investigated in the framework of a quantum field theory of the Dirac
equation, however a simpler and more intuitive approach is possible
within the one-particle Dirac equation (7), where pair production
corresponds to a field-induced transition of an electron in the
Dirac sea of negative-energy states to a final positive-energy state
\cite{oscillating2,oscillating3}. In this approach, a simple picture
of pair production as a wave packet break-up process has been
suggested \cite{oscillating3}. Here an electron in the negative sea
is represented by a Gaussian wave packet formed by a superposition
of negative-energy states. When the $e^+e^-$ pair is produced under
the action of the oscillating field, a droplet is separated from the
wave packet and moves opposite to the initial one (see, for
instance, Fig.1 of Ref.\cite{oscillating3}). The droplet is a
positive-energy state and represents the created electron. It has
been shown that such a simplified model of pair production, besides
to be rather intuitive, also provides a correct way to calculate the
rate of particle creation \cite{oscillating3}. Our classical wave
optics analogue, grounded on the formal analogy between the
one-particle Dirac equation and the spinor-like wave equation (7)
via the variable transformation (9), thus enables to visualize in
space the analogue of pair production as a break up of an initial
Gaussian wave packet, composed by a superposition of Bloch modes of
the lowest lattice miniband and representing an electron in the
Dirac sea, under the influence of an oscillating electric field.

\subsection{Two-level model}
For a spatially-homogeneous and time-dependent field, it is known
that momentum conservation reduces the problem of pair creation to a
two-level system consisting of a negative and a positive energy
state coupled by the external field. Within the two-level model,
pair production generally occurs as a multiphoton resonance process
enforced by energy conservation, with interesting effects such as
Rabi oscillations of the quantum vacuum (see, for instance,
\cite{oscillating2,oscillating3}). The two-level description of pair
production in the dipole approximation for the one-particle
 Dirac equation can be found, for instance, in
Refs.\cite{oscillating2}; for the sake of completeness, it is
briefly reviewed in the Appendix for the case of the one-dimensional
Dirac equation. Owing to the quantum-optical analogy established in
Sec.II.B, the same dynamical scenario is thus expected for light
transport in a periodically-curved tight-binding binary array. To
this aim, let us look for a solution to coupled-mode equations (5)
of the form
\begin{equation}
a_l(z)= \left(
\begin{array}{c}
s_1(z) \\
s_2(z)
\end{array}
\right) \exp(iqla)
\end{equation}
\begin{figure}[htbp]
\includegraphics[width=85mm]{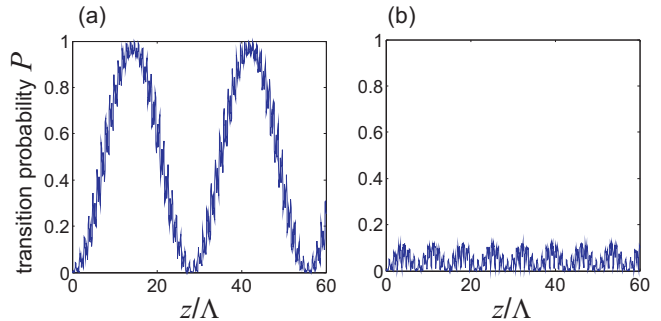}\\
   \caption{(color online) Transition probability $P=|r_+(z)|^2$ versus propagation distance $z$, normalized to the ac oscillation period $\Lambda$,
   in a tight-binding binary lattice with a sinusoidally-curved axis
   for parameter values $\sigma=2$, $\delta=1.817$, $qa=\pi/4$, $\Phi_0=0.4$ and for two
   values of modulation period $\Lambda$: (a) $\Lambda=2.8556$, and (b) $\Lambda=2.7196$. In (a) the multiphoton resonance condition (22)
   is satisfied for $n=3$, and Rabi flopping is clearly observed. In (b) the modulation period is slightly decreased by $5\%$ from
   the $n=3$ resonance value.}
\end{figure}
where the wave number $q$ is related to the electron momentum $k$
for the Dirac equation (7) by Eq.(11), and where the upper (lower)
row applies to an even (odd) value of index $l$. Substitution of
Eq.(12) into Eq.(5) yields the following coupled equations
\begin{eqnarray}
         i \frac{d}{dz} \left(
         \begin{array}{c}
         s_1 \\
         s_2
         \end{array}
         \right)=\mathcal{T}(z) \left(
         \begin{array}{c}
         s_1 \\
         s_2
         \end{array}
         \right)
\end{eqnarray}
where
\begin{equation}
\mathcal{T}(z)= \left(
         \begin{array}{cc}
         \delta & -2 \sigma \cos[qa-\Phi(z)] \\
         -2 \sigma \cos[qa-\Phi(z)] & -\delta
         \end{array}
         \right).
\end{equation}
For a straight array, i.e. in absence of the external field
$\Phi=0$, the (normalized) eigenvectors of the $z$-independent
matrix $\mathcal{T}$,
\begin{equation}
\left(
         \begin{array}{c}
         s_1 \\
         s_2
         \end{array}
 \right)_{ \pm} =
 \left(
         \begin{array}{c}
         -\frac{2 \sigma \cos(qa)}{\sqrt{2 |\omega_{\pm}(q)| |\omega_{\pm}(q)-\delta|}} \\
\frac{\omega_{\pm}(q)-\delta}{\sqrt{2 | \omega_{\pm}(q)|
|\omega_{\pm}(q)-\delta|}}
         \end{array}
 \right),
\end{equation}
corresponding to the eigenvalues $\omega_{\pm}(q)$ defined by
Eq.(3), belong to the positive (electron) and negative (positron)
energy branches. These states are basically the Bloch modes of the
superlattice, belonging to the first two lowest bands, with wave
number $q$. If the initial state
 at $z=0$ is prepared in the Bloch
mode belonging to the negative energy branch (corresponding to an
electron occupying a negative energy state of the Dirac sea in the
quantum mechanical context), i.e. if $(s_1,s_2)^T=(s_1,s_2)_{-}^{T}$
at $z=0$, the application of the oscillating external field
$\Phi(z)$ can induce a transition to the
 Bloch mode $(s_1,s_2)_{+}^{T}$ on the positive energy branch, i.e.
the optical analogue of an electron-positron pair production is
realized. It should be noted that direct (and even indirect)
interband transitions and Rabi oscillations induced by a suitable
longitudinal refractive index modulation in singly-periodic
waveguide arrays have been recently predicted and experimentally
observed in Refs.\cite{SegevRabi}, however in those previous works
the transitions are not described
in terms of a Dirac-type equation with an oscillating field, and the
quantum-optical analogy is thus not evident \cite{note2}.\\
To study the field-induced transition process, it is worth
projecting the state $(s_1,s_2)$ onto the Bloch mode basis (15),
i.e. let us set
\begin{equation}
\left(
         \begin{array}{c}
         s_1 \\
         s_2
         \end{array}
 \right)=r_-(z) \left(
         \begin{array}{c}
         s_1 \\
         s_2
         \end{array}
 \right)_- +
r_+(z) \left(
         \begin{array}{c}
         s_1 \\
         s_2
         \end{array}
 \right)_+
\end{equation}
where $r_-(z)$ and $r_+(z)$ are the occupation amplitudes of the
negative and positive electron branches, respectively. Substitution
Eq.(16) into Eq.(13) yields the following coupled equations for the
occupation amplitudes
\begin{eqnarray}
         i \frac{d}{dz} \left(
         \begin{array}{c}
         r_- \\
         r_+
         \end{array}
         \right)=\mathcal{Z}(z) \left(
         \begin{array}{c}
         r_- \\
         r_+
         \end{array}
         \right)
\end{eqnarray}
where the $z$-varying elements of the $2 \times 2$ matrix
$\mathcal{Z}(z)$ are given by
\begin{eqnarray}
\mathcal{Z}_{11} & = & -\mathcal{Z}_{22}  =  \frac{\delta^2+4
\sigma^2 \cos
(qa) \cos[qa-\Phi(z)]}{\omega_+(q)} \\
\mathcal{Z}_{12} & = & \mathcal{Z}_{21}  =  \frac{2 \sigma \delta
\left\{ \cos(qa)-\cos[qa-\Phi(z)]\right\} }{\omega_+(q)} .
\end{eqnarray}
Equations (17) are generally encountered in problems of driven
two-level systems, and therefore phenomena like population Rabi
flopping, multiphoton resonances, etc. are expected to occur in our
system as well. Note that the conservation of population implies
that $|r_-(z)|^2+|r_+(z)|^2=1$. For the initial condition $r_-(0)=1$
and assuming that the external ac field is switched on at $z=0$ and
off after a propagation distance $z=L_0$ (which is assumed to be an
integer multiple of the modulation cycle $\Lambda$), i.e.
$\Phi(z)=0$ for $z>L_0$, the transition probability is thus given by
$P=|r_{+}(L_0)|^2$. The connection between the two-level model (17),
derived from the tight-binding lattice model (5), and the two-level
model of pair production in the framework of the one-dimensional
Dirac equation (7) is discussed in the Appendix. The latter model is
retrieved from the former one by assuming a wave number $q$ near the
edge of the Brilluoin zone [Eq.(11) with $|k| \ll \pi/2$] and a
small value of $|\Phi(z)|$; in this case, the expressions of matrix
coefficients take the following simplified form [see also Eqs.(A9)
and (A10) given in the Appendix]
\begin{eqnarray}
\mathcal{Z}_{11} & = & -\mathcal{Z}_{22}  = \epsilon(k)-\frac{2 \sigma^2 k \Phi(z)}{\epsilon(k)} \\
\mathcal{Z}_{12} & = & \mathcal{Z}_{21}  = -\frac{2 \sigma \delta
\Phi(z)}{\epsilon(k)}
\end{eqnarray}
where $\epsilon(k)$ is defined by Eq.(10). It is worth noticing
that, under such an assumption, Eqs.(17) are analogous to the
two-level equations describing population dynamics in a two-level
dipolar molecule under an external ac field (see, for instance,
\cite{Longhi06JPB}).\\
In the context of field-induced pair production, the photon energy
$2 \pi \hbar/ \Lambda$ of the ac field is generally much smaller
that the energy separation $2 mc^2$ between negative and positive
energy states, and the transition of one electron out of the Dirac
sea into a positive-energy state generally occurs via a multiphoton
resonant process (see, for instance,
\cite{oscillating2,oscillating3}). The condition of $n$-photon
resonance, obtained in the low ac frequency limit by a WKB analysis
of Eqs.(17-19), reads explicitly (see, e.g., \cite{oscillating3})
\begin{equation}
n \frac{2 \pi}{\Lambda}= 2 \mathcal{E}(q),
\end{equation}
where the quasi-energy $\mathcal{E}(q)$ is calculated as
\begin{eqnarray}
\mathcal{E}(q) & = & \frac{1}{\Lambda} \int_0^{\Lambda} dz
\sqrt{Z_{11}^2(z)+Z_{12}^2(z)}  \\
& = & \frac{1}{2 \pi} \int_{0}^{2 \pi} dy \sqrt{\delta^2+4 \sigma^2
\cos^2[qa-\Phi(\Lambda y / 2 \pi)]}. \nonumber
\end{eqnarray}
As an example, Fig.2(a) depicts the behavior of the transition
probability $P=|r_+(z)|^2$ versus $z$, showing multiphoton Rabi
oscillations, as obtained by numerical simulations of Eqs.(17-19)
for a sinusoidal ac field $\Phi(z)=\Phi_0 \sin(2 \pi z / \Lambda)$
for parameter values $\sigma=2$, $\delta=1.817$, $qa=\pi/4$,
$\Phi_0=0.4$ and for a modulation period $\Lambda=2.8556$, which
satisfies the multiphoton resonance condition (22) with $n=3$. For
comparison, Fig.2(b) shows the behavior of the transition
probability for the same parameter values, but for a modulation
period $\Lambda=2.7196$ which is $5\%$ smaller than the resonant
value of Fig.2(a). A comparison of Figs.2(a) and (b) clearly
indicates that, as it is well known,  multiphoton pair production is
a resonant process.\\
In our photonic system, the constraints of low modulation frequency
and low field amplitudes typical of laser-driven QED vacuum can be
overcome. For typical geometrical settings which apply to binary
waveguide arrays and for reasonable propagation lengths achievable
with current samples, the regimes of fast modulation frequencies
(i.e., short bending periods $\Lambda$) and strong amplitudes are
indeed more accessible than the low-frequency multiphoton resonance
regime; on the other hand, such a latter regime might be accessible
for cold atoms in optical superlattices \cite{Breid}. For instance,
the values of $\sigma$ and $\delta$ used in Fig.2 correspond to the
coupling rate and propagation constant mismatch, in units of ${\rm
cm}^{-1}$, of a typical binary waveguide array for parameters values
discussed in the next subsection. For such an array, the modulation
period of axis bending in Fig.2(a) that achieves the $n=3$ photon
resonance condition is $\Lambda=2.8556$ cm, and the multiphoton Rabi
oscillation period shown in Fig.2(a) would thus correspond to about
$80$ cm, a length which is not accessible in an experiment.
Conversely, the regimes of fast modulation frequency and strong
bending amplitudes are easily accessible. In such regimes, efficient
transition can occur even for a single-cycle of the ac field
[$\Phi(z)=\Phi_0 \sin(2 \pi z/ \Lambda)$ for $0<z<\Lambda$,
$\Phi(z)=0$ for $z<0$ and $z>\Lambda$], corresponding to the use of
ultrastrong single-cycle counter-propagating laser pulses in the QED
context. As an example, Fig.3(a) shows the behavior of the
transition probability $P$, after the interaction with a
single-cycle pulse, as obtained by numerical simulations of
Eqs.(17-19) for $\sigma=2$, $\delta=1.817$, $qa=\pi/4$ (as in Fig.2)
but with a much shorter period $\Lambda= 0.6676$, and for increasing
values of the the field amplitude $\Phi_0$. Examples of the detailed
behavior of $P(z)=|r_+(z)|^2$ along the oscillation cycle,
corresponding to $\Phi_0=4$ and $\Phi_0=6$ [points A and B in
Fig.3(a)], are depicted in Figs.3(b) and (c), respectively.
\begin{figure}[htbp]
\includegraphics[width=85mm]{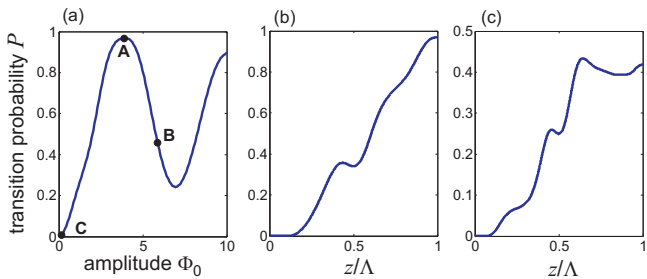}\\
   \caption{(color online) Transition probability $P$ in a tight-binding binary lattice with a single-cycle sinusoidally-curved axis
   for parameter values $\sigma=2$, $\delta=1.817$, $qa=\pi/4$, $\Lambda=0.6676$, and for increasing values of
    amplitude $\Phi_0$. Points A, B and C in the figure
    correspond to the values $\Phi_0=4$, $\Phi_0=6$ and $\Phi_0=0$,
    respectively. (b) and (c) show the detailed behavior of
$|r_+(z)|^2$ versus propagation distance $z$ in the single-cycle
modulation period corresponding to points A [Fig.3(b)] and B
[Fig.3(c)] of Fig.3(a).}
\end{figure}

\subsection{Wave packet dynamics}

The two-level description of pair production in a
spatially-homogeneous and oscillating field described in the
previous subsection assumes states with definite momentum $p$ in the
negative and positive energy branches, and thus fully delocalized in
space. A better visualization of the pair creation process is
attained by considering, as an initial state, a free wave packet in
the negative-energy continuum (for instance Gaussian-shaped),
representing an electron in the Dirac sea \cite{oscillating3}. After
the external ac field is switched on for some interval (for instance
for one oscillation cycle) and then switched off again, transitions
into the positive-energy continuum is visualized as a break up of
the initial wave packet into two wave packets, which propagates with
different group velocities and thus separate in space after some
time \cite{oscillating3}. These two wave packets represent the
amplitude probabilities for the electron to be excited in the
positive-energy continuum or to remain in the Dirac sea. In our
photonic analogue, wave packet break up is simply explained because
of the different refraction angles of wave packets belonging to the
two minibands of the superlattices: the creation of a wave packet in
the upper lattice miniband, induced by the longitudinal bending of
the waveguide axis, is simply observed as a deviation of the
propagation direction from that of the initial wave packet.\\
We have checked such a scenario by direct numerical simulations of
the paraxial field equation (1) using a standard pseudospectral
split-step method.  Parameter values and refractive index profiles
used in the simulations are compatible with binary wavegude arrays
realized in fused silica by femtosecond laser writing and excited in
the visible at $\lambda=633$ nm \cite{Dreisow09}.
\begin{figure}[htbp]
\includegraphics[width=85mm]{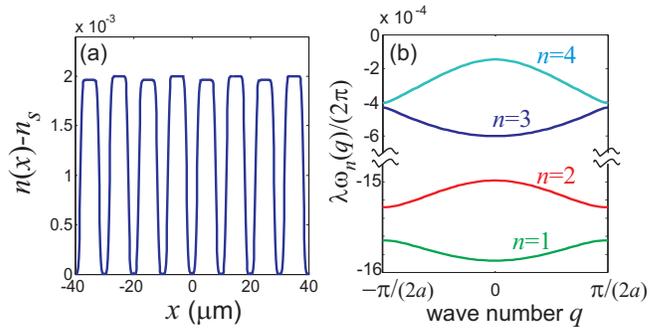}\\
   \caption{(color online) (a) Refractive index profile of the binary superlattice used in numerical simulations, and (b) corresponding band
    diagram. Parameter values are given in the text.}
\end{figure}
Figure 4(a) shows the index profile $n(x)$ of the superlattice used
in the simulations, corresponding to a waveguide spacing $a=10 \;
\mu$m, refractive index changes $dn_1=0.002$ and $dn_2=0.00196$ of
adjacent waveguides, and a substrate refractive index $n_s=1.42$.
Figure 4(b) shows the dispersion curves $\omega_n(q)$ of a few
low-order bands of the straight superlattice (band diagram).
$\omega_n(q)$ are computed by a standard plane-wave expansion method
by looking for a solution to Eq.(1), with $F=0$, of Bloch-Floquet
type, i.e. of the form $E(x,z)=u_n(x,q) \exp(iq x) \exp[-i
\omega_n(q)z]$, where $n$ is the band order, $q$ varies in the first
Brillouin zone $-\pi/(2a) < q \leq \pi/(2a)$, and $u_n(x,q)$ in the
periodic part of the Bloch mode [$u_n(x+2a,q)=u_n(x,q)$]. Note that
the two lowest bands $n=1$ and $n=2$ in Fig.4(b) correspond to the
two minibands $\omega_-$ and $\omega_+$, respectively, of the
tight-binding model (2) depicted in Fig.1(c), i.e. to the negative-
and positive-energy branches of the Dirac equation (7) in absence of
the external field. For the binary array of Fig.4, the coupling rate
$\sigma$ and propagation constant mismatch $\delta$ entering in the
tight-binding model (2) can be simply estimated by fitting the two
lowest bands of Fig.4(b) using Eq.(3) with $\sigma$ and $\delta$ as
fitting parameters, yielding $\sigma \simeq 2 \; {\rm cm}^{-1}$ and
$\delta \simeq 1.817 \; {\rm cm}^{-1}$. To visualize the optical
analogue of pair production, the bending profile of the waveguide
axis is designed  to simulate the action of a single-cycle
ultrastrong field in the QED context, namely $x_0(z)=-A \cos(2 \pi
z/ \Lambda)$ for $0<z<\Lambda$, and $x_0(z)=-A$ for $z>\Lambda$. The
array is excited at the input plane by a broad Gaussian beam of spot
size $w_0$, tilted at the angle $\theta$, i.e. Eq.(1) is integrated
with the initial condition $E(x,0)=\exp[-(x/w_0)^2] \exp( 2 \pi i
n_s x \theta / \lambda)$. The tilt angle has been chosen half of the
Bragg angle, i.e. $\theta=\theta_B/2$, where $\theta_B \simeq
0.64^{\rm o}$. At such a relatively small excitation angle, the
lowest band of the array is mostly excited (see, for instance,
\cite{Longhi06OL}), and therefore the Gaussian wave packet mimics an
initial electron wave packet of the Dirac sea, i.e. in the
negative-energy spectrum. For $\theta=\theta_B/2$, the condition
$qa=\pi/4$ of Fig.3 is also satisfied. The modulation period has
been set equal to $\Lambda \simeq 0.67$ cm to reproduce the
conditions of Fig.3.
\begin{figure}[htbp]
\includegraphics[width=70mm]{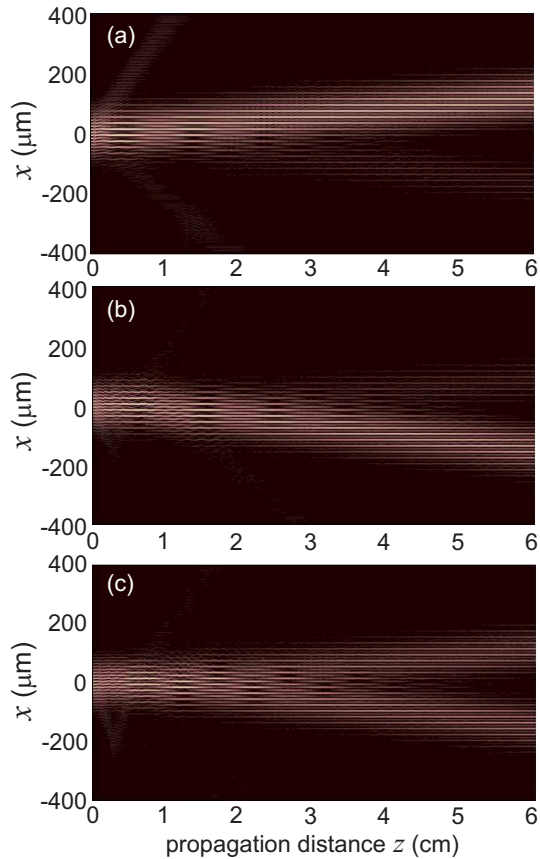}\\
   \caption{(color online) Beam propagation in a 6-cm-long one-dimensional binary
   waveguide array (snapshot of $|E(x,z)|^2$), comprising a first section with a single-cycle sinusoidally curved
   axis of period $\Lambda=6.67 \; {\rm}$ and amplitude $A$, and a second section with a straight
   axis: (a) $A=0$ (straight array), (b) $A=30 \; \mu$m, and (c) $A=45 \; \mu$m
   The array is excited by a Gaussian beam tilted at half of the Bragg angle
   (parameter values are given in the text).}
\end{figure}
Figure 5 shows a few examples of beam propagation along a 6-cm-long
array for an initial beam spot size $w_0=80 \; \mu$m and for
increasing values of the modulation amplitude $A$. In Fig.5(a), the
array is straight [$A=0$, corresponding to point C of Fig.3(a)] and
the wave packet propagates weakly diffracting with a nonvanishing
transverse group velocity along a direction which is determined by
the normal to the band dispersion curve at $qa=\pi/4$. In Fig.5(b),
the amplitude of the modulation is increased to $A=30 \; \mu$m,
corresponding to $\Phi_0 \simeq 4$ [see Eq.(6)], i.e. to point A of
Fig.3(a). In this case, after the ac field is switched off at
$z=\Lambda \simeq 6.7$ mm, the wave packet does not basically break
down, however it is now refracted in an opposite direction as
compared to the case of Fig.5(a). This is the clear signature that
the wave packet is mostly composed by Bloch modes of the second
band, i.e. that an almost complete transition from negative- to
positive-energy states has occurred, according to the predictions of
the two-level analysis [see point A of Fig.3(a)]. In Fig.5(c), the
amplitude of the modulation is further increased to $A=45 \; \mu$m,
corresponding to $\Phi_0 \simeq 6$. According to the prediction of
the two-level analysis [see point B of Fig.3(a)], after the ac field
is switched off the transition probability for the wave packet
components is $\sim 45 \%$. Accordingly, the wave packet breaks up
at propagation distances $z>\Lambda$, with the two nearly balanced
wave packets belonging to the first and second array minibands
refracting along two opposite directions.

\section{Conclusions}
Vacuum decay and electron-positron pair production due to the
instability of the QED vacuum in an external oscillating field are
remarkable predictions of the Dirac theory. Their experimental
demonstration, however, requires ultrastrong and high-frequency
laser fields which are not yet available. In this work a classic
wave optics analogue of vacuum decay has been proposed which is
based on the formal analogy between the temporal dynamics of the
one-dimensional Dirac equation for a single particle in an external
spatially-homogeneous oscillating field and the spatial propagation
of monochromatic light waves in a periodically-curved binary
waveguide array. Our photonic analogue enables a simple and
experimentally-accessible visualization in space of the process of
pair production as break up of an initially negative-energy Gaussian
wave packet, representing an electron in the Dirac sea, under the
influence of an oscillating electric field.

\appendix

\section{Two-level description of pair production in the dipole approximation}
In this Appendix we briefly review the two-level description of pair
production, induced by an oscillating field in the dipole
approximation, for the one-dimensional single-particle Dirac
equation given by Eq.(7) in the text. In making such an analysis, we
introduce the variable transformation defined by Eqs.(9), and
re-write Eq.(7) into the more familiar quantum mechanical form
\begin{equation}
i \hbar \frac{\partial \psi}{\partial t}=(cp_x-eA_x) \alpha
\psi+mc^2 \beta \psi,
\end{equation}
where $p_x=-i \hbar \partial_x$. The independence of the vectorial
potential $A_x$ on the spatial coordinate $x$ corresponds to the
electric dipole approximation and to the interaction with a
spatially-homogeneous oscillating  electric field. Owing to the pure
time dependence of the external potential, the momentum is conserved
during the transition. Therefore only transitions between negative-
and positive-energy states with the same momentum are permitted in
the pair creation process. Indicating by $p=\hbar k$ the momentum of
the particle at initial time, a solution to Eq.(7) with definite
momentum is given by
\begin{equation}
\psi(x,t)=\varphi(t) \exp(ikx)
\end{equation}
where the two-component spinor $\varphi=(\varphi_1,\varphi_2)^T$
satisfies the equation
\begin{equation}
i \frac{d \varphi}{dt}=\mathcal{M}(t) \varphi.
\end{equation}
The time-dependent elements of the $2 \times 2$ matrix $\mathcal{M}$
in Eq.(A3) are given by
\begin{eqnarray}
\mathcal{M}_{11} & = & -\mathcal{M}_{22}=\frac{mc^2}{\hbar} \\
\mathcal{M}_{12} & = & \mathcal{M}_{21}=kc-\frac{eA_x(t)}{\hbar}.
\end{eqnarray}
To investigate field-induced transitions, let us project the spinor
$\varphi(t)$ on the basis of the two eigenstates $\varphi_{\pm}$ of
the free electron (i.e., in the absence of the external field)
corresponding to the negative and positive energy branches, i.e. let
us set
\begin{equation}
\varphi(t)=r_-(t) \varphi_- +r_+(t) \varphi_+
\end{equation}
where
\begin{eqnarray}
\varphi_{\pm}=\frac{1}{\sqrt{p^2c^2+[\hbar \epsilon(p) \mp mc^2
]^2}} \left(
\begin{array}{c}
cp \\
\pm \hbar \epsilon(p)-mc^2
\end{array}
\right),
\end{eqnarray}
$ \hbar \epsilon(p)=\sqrt{p^2c^2+(mc^2)^2}$, and $r_{\pm}(t)$ are
the occupation amplitudes of negative and positive energy states at
time $t$. The amplitudes $r_-(t)$ and $r_+(t)$ then satisfy the
coupled equations of driven two-level systems
\begin{equation}
i \frac{d}{dt} \left(
\begin{array}{c}
r_- \\
r_+
\end{array}
\right)= \mathcal{Z}(t) \left(
\begin{array}{c}
r_- \\
r_+
\end{array}
\right)
\end{equation}
with matrix elements given by
\begin{eqnarray}
\mathcal{Z}_{11} & = & -\mathcal{Z}_{22}  =  \epsilon(p)-\frac{pceA_x(t)}{\hbar^2 \epsilon(p)} \\
\mathcal{Z}_{12} & = & \mathcal{Z}_{21}  = -\frac{mc^2
eA_x(t)}{\hbar^2 \epsilon(p)}.
\end{eqnarray}
Equations (A8-A10) are analogous to Eqs.(17-19) given in text for
the driven tight-binding lattice (5). In particular, according to
the approximations used to derive the Dirac equation (7) in
Sec.II.B, the expressions of the matrix coefficients
$\mathcal{Z}_{ik}(t)$, defined by Eqs.(18) and (19), reduce to those
given by Eqs.(A9) and (A10), after the replacement $qa \rightarrow
\pi/2+k/2$ [see Eq.(11)], assuming $|k| \ll \pi/2$, and using the
transformation of variables (9) that connect the quantum and optical
descriptions.

%\clearpage
%\bibliography{H:/Physik/bibliography}

\begin{thebibliography}{31}
\bibitem{note1}
Experimental observations of the vacuum decay and corresponding
$e^+e^-$ pairs creation triggered solely by pure laser light are
still lacking, though they may be envisaged in the near future with
the development of new ultrastrong laser system facilities.
\end{thebibliography}

\end{document}